\documentclass[11pt,reqno]{amsart}
\usepackage{fullpage} % Package to use full page
\usepackage{parskip} % Package to tweak paragraph skipping
\usepackage{amsmath,amssymb,amsfonts,amsthm}
\usepackage{mathrsfs}
\usepackage{graphicx,textcomp}

\usepackage[cmtip,all]{xy}

\usepackage{TqcCommands}%ZW package
\usepackage{tikz, braids} % For graphics, drawing braids
\usetikzlibrary{knots,hobby}

\newtheorem{theorem}{Theorem}[section]

\theoremstyle{definition}
\newtheorem{definition}{Definition}[section]

\theoremstyle{remark}

\theoremstyle{proposition}

\theoremstyle{conjecture}
\newtheorem{conjecture}{Conjecture}[section]

\numberwithin{equation}{section}
\DeclareMathOperator{\End}{End}

\DeclareMathOperator{\tr}{tr}

\begin{document}

\title{Braids, Motions and Topological Quantum Computing}


%    author one information
\author{Eric C. Rowell}
\address{Department of Mathematics\\
    Texas A\&M University \\
    College Station, TX 77843\\
    U.S.A.}
\curraddr{}
\email{rowell@math.tamu.edu}
\thanks{  Rowell is partially supported by NSF grants DMS-1664359 and DMS-2000331, and thanks Zhenghan Wang, Paul Martin and Xingshan Cui for helpful conversations.}

%    author two information

\date{}

\dedicatory{}

%    Abstract is required.

\begin{abstract} The topological model for quantum computation is an inherently fault-tolerant model built on anyons in topological phases of matter.  A key role is played by the braid group, and in this survey we focus on a selection of ways that the mathematical study of braids is crucial for the theory. We provide some brief historical context as well, emphasizing ways that braiding appears in physical contexts. 
 We also briefly discuss the 3-dimensional generalization of braiding: motions of knots.\end{abstract}

\maketitle


\section{Introduction}
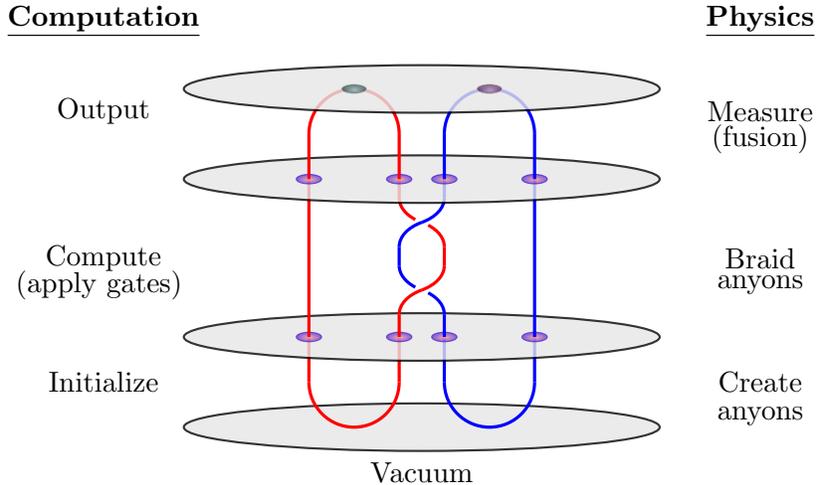
\begin{figure}[ht]\begin{tikzpicture}[xscale=0.3,yscale=0.3]
\filldraw[fill=white!90!black,draw=black,opacity=0.8,thick] (-1,-4) ellipse (300pt and 30 pt);
  % creation anyons
  \path[draw=red, very thick] (-2,-2) -- (-2,0);
  \path[draw=red, very thick] (-6,-2) -- (-6,0);
  \path[draw=red, very thick] (-6,-2) arc (180:360:2 and 2);
  \path[draw=blue, very thick] (0,-2) -- (0,0);
  \path[draw=blue, very thick] (4,-2) -- (4,0);
  \path[draw=blue, very thick] (0,-2) arc (180:360:2 and 2);
  %before braiding layer
\filldraw[fill=white!90!black,draw=black,opacity=0.8,thick] (-1,0) ellipse (300pt and 30 pt);
  % quasi-particles before braiding
  \draw[rotate=0,inner color=red!50!, outer color=red!20!blue, draw=red!20!blue,opacity=0.6] (0,0) ellipse (16pt and 6pt);
  \draw[rotate=0,inner color=red!50!, outer color=red!20!blue, draw=red!20!blue,opacity=0.6] (4,0) ellipse (16pt and 6pt);
  \draw[rotate=0,inner color=red!50!, outer color=red!20!blue, draw=red!20!blue,opacity=0.6] (-2,0) ellipse (16pt and 6pt);
  \draw[rotate=0,inner color=red!50!, outer color=red!20!blue, draw=red!20!blue,opacity=0.6] (-6,0) ellipse (16pt and 6pt);
  % Vacuum start
  \node[] () at (-1,-6) {Vacuum};
  % initialize/particle creation
  \node[] () at (-15.1, -2) {Initialize};
  \node[] () at (14, -2) {Create};% particles};
  \node[] () at (14, -3.5) {anyons};% particles};
  % computation/physics header
  \node[] () at (-15.1,14) {\textbf{\underline{Computation}}};
  \node[] () at (14,14) {\textbf{\underline{Physics}}};
  \path[draw=red, very thick] (-2,0)--(-2,1);
  \path[draw=blue, very thick] (0,0)--(0,1);
  \path[draw=red, very thick] (-2,1) arc (-180:-240:2 and 1.25);
  \path[draw=red, very thick] (-1,2.08253) arc (300:360:2 and 1.25);
  \path[draw=blue, very thick] (0,1) arc (0:50:2 and 1.25);
  \path[draw=blue, very thick] (-2,3.16506) arc (180:230:2 and 1.25);
  \path[draw=blue, very thick] (-2,3.16506) -- (-2,4);
  \path[draw=red, very thick] (0,3.16506) -- (0,4);

  \path[draw=blue, very thick] (-2,4) arc (-180:-240:2 and 1.25);

  \path[draw=blue, very thick] (-1,5.08253) arc (300:360:2 and 1.25);

  \path[draw=red, very thick] (0,4) arc (0:50:2 and 1.25);

  \path[draw=red, very thick] (-2,6.16506) arc (180:230:2 and 1.25);
  \path[draw=red, very thick] (-2,6.16506) -- (-2,7);
  \path[draw=blue, very thick] (0,6.16506) -- (0,7);
  \path[draw=red, very thick] (-6,0) -- (-6, 7);
  \path[draw=blue, very thick] (4,0) -- (4,7);
\filldraw[fill=white!90!black,draw=black,opacity=0.8, thick] (-1,7) ellipse (300pt and 30 pt);
  \draw[rotate=0,inner color=red!50!, outer color=red!20!blue, draw=red!20!blue,opacity=0.6] (0,7) ellipse (16pt and 6pt);
  \draw[rotate=0,inner color=red!50!, outer color=red!20!blue, draw=red!20!blue,opacity=0.6] (4,7) ellipse (16pt and 6pt);

  \draw[rotate=0,inner color=red!50!, outer color=red!20!blue, draw=red!20!blue,opacity=0.6] (-2,7) ellipse (16pt and 6pt);

  \draw[rotate=0,inner color=red!50!, outer color=red!20!blue, draw=red!20!blue,opacity=0.6] (-6,7) ellipse (16pt and 6pt);

  % compute/braid

  \node[] () at (-15.1, 3.5) {Compute};

  \node[] () at (-15.3, 2.3) {(apply gates)};

  \node[] () at (14, 3.5) {Braid};%Particle exchange (braid)};

\node[] () at (14, 2.3) {anyons};

  \path[draw=red, very thick] (-6, 7) -- (-6, 9);

  \path[draw=red, very thick] (-2,7) -- (-2, 9);

  \path[draw=red, very thick] (-2,9) arc (0:180:2 and 2);

  \path[draw=blue, very thick] (0, 7) -- (0, 9);

  \path[draw=blue, very thick] (4,7) -- (4, 9);

  \path[draw=blue, very thick] (4,9) arc (0:180:2 and 2);
\filldraw[fill=white!90!black,draw=black,opacity=0.8, thick] (-1,11) ellipse (300pt and 30 pt);

  % quasi particles after fusion

  \draw[rotate=0,inner color=teal!50!, outer color=teal!20!black, draw=teal!20!,opacity=0.6] (-4,11) ellipse (16pt and 6pt);

  \draw[rotate=0,inner color=violet!50!, outer color=violet!20!black, draw=violet!20!,opacity=0.6] (2,11) ellipse (16pt and 6pt);

  % output/measure

  \node[] () at (-15.1, 10) {Output};

  \node[] () at (14, 10) {Measure };

\node[] () at (14, 8.8) {(fusion)};

  % vacuum at end?

  \node[] () at (-1,14){};% {Vacuum};

\end{tikzpicture}
 \caption{\label{fig:TQC}A single iteration of a topological quantum computation via braiding.}

\end{figure}

Quantum computation is predicated upon the ability to create, manipulate and measure the states in quantum systems \cite{FKLW03}. In the topological model for quantum computation 
 \cite{FKLW03,NSSFD08} the quantum systems of interest are topological phases of matter. Typically, these are 2 dimensional systems harboring point-like excitations--anyons \cite{wil90}. This alternative to the more traditional quantum circuit model is theoretically robust against decoherence due to the topological nature of the of the states: the information is stable due to invariance of the states under small, local perturbations.  On the other hand, in order to implement quantum gates more drastic evolution of the states must be employed.  The most natural choice is particle exchange of the anyons, so that their trajectories in $2+1$ dimensional space-time form braids.  The corresponding transformations are the quantum gates, which are mathematically encoded in unitary representations of the braid group.  The process of creating anyons, braiding them and then measuring form knots or links in (2+1)-dimensional spacetime, which is interpreted as a (single run) of a topological quantum computation as illustrated in figure \ref{fig:TQC}. 
Notice that the motions of the point-like excitations in the effectively two dimensional medium have a natural mathematical interpretation as the group of braids.
Thus it comes as no surprise that the braid group $\mathcal{B}_n$ plays an outsized role in the development of topological quantum computation.
In this article we will describe a selection of topics illustrating the connection between braids and topological quantum computation as well as some generalizations.  We make no attempt to be exhaustive on the subject, nor will we provide a 
completely self-contained background.  There are a number of much more complete treatments of topological quantum computation from various perspectives and for various tastes, a few of which we list here: \cite{Wang06,Wang13,DRW16,NSSFD08,Wa10,Pa12,OP99,FKLW03,Rowell16,RWBull}.  For deeper and more complete mathematical details on the braid group and invariants of knots and links we suggest \cite{KasTur}, while for the categorical background the text \cite{EGNOBook} is an excellent resource.

\subsection{The Braid Group}
The braid group $\mcB_n$ on $n$ strands is most concisely described using the abstract group presentation by generators and relations due to Artin \cite{Artin25}: 
\begin{align*}\mcB_n =  \langle \sigma_1, \sigma_2, \ldots, \sigma_{n-1} \mid &\sigma_i\sigma_j=\sigma_j \sigma_i \text{ for } |i -j| \ge 2,\\ &\sigma_i\sigma_{i+1}\sigma_i=\sigma_{i+1}\sigma_i\sigma_{i+1}, \text{ for }  1\leq i,j \leq n-1  \rangle\end{align*}

The first relation is referred to as far commutativity and the second is the braid relation.

 For visualization purposes we display the generator $\sigma_1$ and its inverse $\sigma_1^{-1}$ geometrically as members of $\mcB_4$:

\[\raisebox{.3cm}{$\sigma_1=\quad$ }\begin{tikzpicture}[scale=.5]
\braid[number of strands=4] (braid)  a_1^{-1}  ;
\end{tikzpicture}\hspace{1cm}\raisebox{.3cm}{$\sigma_1^{-1}=\quad$ }\begin{tikzpicture}[scale=.5]
\braid[number of strands=4] (braid)  a_1  ;
\end{tikzpicture} \]
The composition in the group then corresponds to stacking, whereas the identity is clearly represented by unbraided strands.

Predating Artin's work on braids by 30  years is the work of Hurwitz \cite{Hurwitz91} on motions of points on the 2-dimensional sphere.  This is much closer to their use in condensed matter physics: the world-lines of anyons under particle exchange can be viewed as trajectories of motions of points in the disk $D^2$.  Connecting braids to knots and links are two key mathematical results: Alexander's theorem \cite{Alexander23} and Markov's theorem \cite{Markov36}.  The key consequence of Alexander's theorem is that any knot/link may be obtained the plat closure:  capping off the strands of a braid in $\mcB_{2n}$ pairwise, on both the top and bottom.  Birman's \cite{Birman76} analogue of Markov's theorem establishes a way to determine if two distinct braids have the same plat closure.\footnote{Both Alexander's and Markov's theorems actually deal with a different closure in which one glues the ends of top strands to the the bottom strands.  But the plat closure is more physically relevant.}

\subsection{Braids and Knots in Physics: An Incomplete History}
Around the 7th century B.C.E. there was a debate between the natural philosopher G\=arg\={\i} V\=achaknav\={\i} (daughter of Vachaknu) and the sage Yajnavalkya \cite{Olivelle98}.  She asks him:
\begin{quote}
    Since this whole world is woven back and forth on water, on what then is it woven back and forth?
\end{quote}
To which he replies ``On air, G\=arg\=\i."  The exchange continues with G\=arg\={\i}  asking what air and the intermediate regions are woven upon, and eventually what the universe is woven upon.  This could be the most ancient reference to braids in physics. That the universe consists of a sequence of braidings may seem quaint by today's scientific standards, but the similitude with topological quantum circuits is nonetheless intriguing.\footnote{We thank Paul Martin for informing us of this ancient reference.}

Many point to Gauss as the first person to study knots mathematically and suggest he was inspired by physical considerations such as magnetic potential, see eg., \cite{Epple98,przy98,Ricca11}.  Maxwell himself \cite{Maxwell73} mentions Gauss' integral formula for the linking number of a two component link and its physical interpretation in electromagnetic induction.\footnote{There is some historical controversy here, which the reader may find in the references.}
Gauss' interest in braids goes back even earlier according to \cite{Epple98}, as he drew the following braid sometime between 1815 and 1830:
$$\begin{tikzpicture}[scale=.5]
 \braid[number of strands=4] (braid)  a_1 a_3 a_2^{-1} a_2^{-1}  a_3 ;
\end{tikzpicture} $$
One might naturally surmise that Gauss' early interest in braids and knots were influential in his discovery of the linking number.

Yet another connection between physics and knots came some years later (see \cite{Silver14} for a full account) when Thompson (Lord Kelvin) learned of some experiments of Helmholtz in the 1850s producing knotted vortices of smoke.  This led to the speculation by Thompson that matter was made up of knotted loops of luminiferous \ae ther. This short-lived "vortex-atoms" theory suggested that different elements were distinguished by their knot-type.   Yet again, we find an archaic theory resonating with a modern idea: in topological quantum computation information is stored in knotted world-lines of anyons.  Mathematics reaped the benefit, as Tait produced the first reasonably comprehensive table of knots with few crossings, inspired by his friend Thompson's vortex-atom theory.
\section{Topological Quantum Computation}
Here we provide a historical perspective on topological quantum computation and set up the mathematical framework in terms of categories.
\subsection{The origins of Topological Quantum Computation}
Topological quantum computation emerged as a confluence of ideas of Freedman, in topology, and Kitaev, in physics, \cite{Fr98,Ki03} in the late 1990s.  The framework was then laid out quickly in a series of papers of Freedman, Kitaev, Wang and Larsen \cite{FLW02,FKW02,FKLW03}. 

Freedman postulated that physical systems described by  topological quantum field theories (TQFTs) \cite{Wi89} could potentially be employed to perform certain Jones polynomial \cite{Jo85} evaluations.  
The Jones polynomial is a remarkably powerful invariant of knots and links. Its discovery in the early 1980s \cite{Jo85} precipitated the field of quantum topology. Although it is straightforward to compute for a given knot, the complexity of the competition grows quickly with the number of crossings. Indeed, evaluating the Jones polynomial at roots of unity is classically a \#P-hard computation at most values \cite{JVW90}. A clue to how physics could be useful for such computations is found in the original formulation of the Jones polynomial as the trace of the braid group representation associated with the $SU(2)$ Chern-Simons/Reshetikhin-Turaev TQFTs.

 Meanwhile, Kitaev was interested in topological phases of matter harboring anyons for their potential use in fault-tolerant quantum computation. Topological phases of matter were studied going back to the 1970s and 1980s by the recipients of the 2016 Nobel Prize in Physics \cite{nobel}: Kosterlitz, Thouless and Haldane.  Well-studied examples of include the fractional quantum Hall liquids \cite{MR91,We91} that are expected to yield non-abelian anyons. Further examples include topological insulators \cite{NSSFD08,Majorana15}. More recently nanowires have been studied for their potential for harboring anyons \cite{Mou12,Marcus16}.  The need for expensive error-correction due to decoherence continues to be a major hurdle for scaled up quantum computation.  Kitaev's idea was that quantum gates could be implemented by braiding anyons, so that error-correction comes from the topological nature of the systems.  
 
 Remarkably, Freedman's and Kitaev's proposals are two sides of the same coin: 2D topological phases of matter are modeled by TQFTs!  It is a rare instance when two essentially independent ideas converge yielding a powerful new paradigm.

 To breathe life into this new paradigm some deep mathematics was needed from both quantum topology and group theory. The resulting proof-of-principle of topological quantum computation is found in \cite{FLW02,FKW02}. There they show that 1) the topological model for quantum computation could be as powerful as the more standard quantum circuit model based on qubits and 2) it is not more powerful than the quantum circuit model. The second result is proved by showing that any topological quantum computation can be simulated on a universal qubit machine with polynomial overhead.  The first result show that universal topological quantum computers exist, in principle.  We will return to this below.

\subsection{Modeling Topological Phases with Categories}

Anyons are topological quantum fields emerging as finite energy particle-like excitations in topological phases of matter. Like particles, they can be moved, but cannot be created or destroyed locally. To model them we consider the fusion and braiding structures of these elementary excitations in the plane.  The anyon system is then modeled by a \emph{unitary modular category}.   For this reason we will use \emph{anyon model} and \emph{unitary modular category} synonomously.

There is a philosophical explanation for why category theory is suitable for describing some quantum physics.  In quantum physics we appeal to measurements to ``see" the elementary particles by analyzing their responses to measuring devices.  Anyons are \emph{defined} by how they interact with other particles, and their responses to measuring devices. According to Kapranov and Voevodsky \cite{KV}, the main principle in category theory is: \lq\lq In any category it is unnatural
and undesirable to speak about equality of two objects".  In category theory an object $X$ is determined by the vector spaces of morphisms $\Hom(X,Y)$ for all $Y$ in the tensor category.  Therefore, it is natural to treat objects as  anyons, and the morphisms as models of the quantum processes between them.
For a more complete treatment see the survey \cite{RWBull}.\footnote{We shall not need a full treatment here, we will introduce the key ideas as they become necessary.} The notion of a modular tensor category was invented by Moore and Seiberg using tensors \cite{MS89}, and its coordinate-free version: \emph{modular category} was defined by Turaev \cite{Turaev92}. In this model, an anyon $X$ is a simple object in the category, while the vacuum is regarded as the monoidal unit $\mathbf{1}$.  The Hilbert space of states of a system of anyons on a disk with prescribed boundary condition is modeled by the space of morphisms in the category so that the braiding of anyons provides unitary operators as the system evolves: the quantum gates in topological quantum computation.  For example, the Hilbert space associated with $n$ indistinguishable $X$ anyons in the disk with boundary condition (total charge) $Y$ is the space $\Hom(Y,X^{\otimes n})$. One assumption is that there are finitely many distinguishable indecomposable anyon types (including the vacuum type): $\{X_0=\mathbf{1},X_1,\ldots, X_r\}$, so that that every anyon corresponds to some fixed type $a\in\{0,\ldots,r\}$.  A dictionary of terminologies between categories and anyons systems is given in Table~\ref{table-dictionary}, adapted from \cite{Wa10}.  The  Hilbert space $V_{ab}^c=\textrm{Hom}(a\otimes b,c)$ represents the span of the linearly independent ways of fusing anyons of type $a$ and $b$ to obtain the anyon of type $c$.  In topological phases of matter the \emph{ground state degeneracy} is typically encoded as the dimension of the spaces $\bigoplus_cV_{ab}^c$ where the sum is over all anyon types $c$.  Most often this is consider for $a=b$ so that we can speak of the ground state degeneracy of a given anyon. In the vernacular one says $a$ has non-trivial ground-state degeneracy if $\dim(\bigoplus_c V_{aa}^c)>1$.
\begin{table}[hb]
\begin{tabular}{|l|l|}\hline
\emph{UMC} & \emph{Anyonic system}\\\hline
simple object & anyon\\\hline
label & anyon type or topological charge\\\hline
tensor product & fusion\\\hline
%fusion rules & fusion rules\\\hline
triangular space $V^c_{ab}$ or $V^{ab}_c$ & fusion/splitting space\\\hline
dual & antiparticle\\\hline
birth/death & creation/annihilation\\\hline
braid representation & anyon statistics\\\hline
%nonzero vector in $V(Y)$ & ground state vector\\\hline
%unitary $F$-matrices & recoupling rules\\\hline
%twist $\theta_x = e^{2\pi i s_x}$ & topological spin\\\hline
morphism & physical process or operator\\\hline
braids & anyon trajectories\\\hline
%quantum invariants & topological amplitudes\\ \hline
\end{tabular}
\caption{A dictionary of categorical terms and their interpretations in anyonic systems.\label{table-dictionary}}

\end{table}

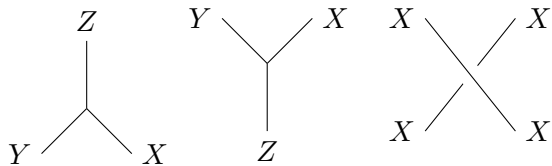
\begin{figure}[th!]
\begin{center}
\begin{tikzpicture}[scale=0.3]%Fig. 2
\begin{scope}[xshift=5cm,yshift=7cm]

\draw (-2,-2)--(0,0);
\draw (2,-2)--(0,0);
\draw (0,0)--(0,3);
 
 \draw (0,3) node[anchor=south] {${Z}$};
 \draw (-2,-2) node[anchor=east] {${Y}$};
 \draw (2,-2) node[anchor=west] {${X}$};
\end{scope}
\begin{scope}[xshift=13cm,yshift=9cm]
 \draw (0,-3)--(0,0);
 \draw (0,0)--(2,2);
 \draw (0,0)--(-2,2);
 \draw (0,-3) node[anchor=north] {${Z}$};
 \draw (-2,2) node[anchor=east] {${Y}$};
 \draw (2,2) node[anchor=west] {${X}$};
\end{scope}
\begin{scope}[xshift=22cm,yshift=9cm]
 \draw (-2,-3)--(2,2);
 \draw [white, line width=3mm] (2,-3)--(-2,2);
 \draw (2,-3)--(-2,2);
 \draw (-2,-3) node[anchor=east] {${X}$};
 \draw (2,-3) node[anchor=west] {${X}$};
  \draw (2,2) node[anchor=west] {${X}$};
 \draw (-2,2) node[anchor=east] {${X}$};
 %\draw (3, -2)--(3,1);
\end{scope}
\end{tikzpicture}

\caption{basic local events: fusion, splitting and braiding\label{fig:elem}}
\end{center}
\end{figure}

There are two basic local events that can occur during a topological quantum computation besides braiding, namely \emph{splitting} and \emph{fusion}. Fusion is when two anyons combine to produce another anyon, while splitting it the reverse.  These events are conveniently represented using the diagrams in Figure \ref{fig:elem}.  A special case of fusion is \emph{annihilation} when an anyon particle and its anti-particle fuse to the vacuum.  Similarly \emph{creation} is a special case of splitting in which a particle/anti-particle pair are drawn from the vacuum. We often depict the splitting/fusion involving the vacuum anyon type $\mathbf{1}$ by a dotted line, or sometimes we omit it entirely.  This is justified as the vacuum anyon may be freely inserted/deleted in the system.  Braiding is the the isomorphism denoted $c_{X,Y}:X\otimes Y\mapsto Y\otimes X$.

As there are finitely many anyon types one may encode the fusion rules into a matrix: for a fixed type $a$, define $[N_{a}]_{c,b}:=\dim(V_{ab}^c)$.  Running over all pairs $(b,c)$ we obtain the fusion matrix $N_a$.  An important invariant of an anyon is the \emph{dimension} $d_a$ of the corresponding anyon type: this is simply the largest eigenvalue of $N_a$, which is guaranteed to be real and positive by the Perron-Frobenius theorem.  However, this dimension is not necessarily an integer--in general it is a real cyclotomic integer: a real number that can be expressed as a finite sum of roots of unity. We shall see later that the computational utility of anyons of type $a$ in a given system is intimately intertwined with the value of $d_a$.  

We are now ready to describe the braid group representations associated with anyons of type $a$ in a topological phase of matter.  Suppose we have $n$ type $a$ anyons localized at positions $1,\ldots, n$ in the system, whose total charge is $b$.  Interchanging the anyons at positions $i$ and $i+1$ induces a unitary operator $U_i(b)$ on the state space $\mathcal{H}(b;a,\ldots,a)=\Hom(b,a^{\otimes n})$, which represents the braid group generator $\sigma_i$.  Extending to all such interchanges and summing over all possible boundary conditions we obtain a representation
\[\rho_{a}:\mcB_n\rightarrow U(\bigoplus_b\mathcal{H}(b;a,\ldots,a))\quad \text{via}\quad \sigma_i\mapsto \bigoplus_b U_i(b).\]
With this formulation in hand, we can mathematically analyze topological gates.  Note that dimension of the representation $\bigoplus_b\mathcal{H}(b;a,\ldots,a)$ is assymptotic to $d_a^n$, so that the ground state degeneracy of $n$ type $a$ anyons in the disk can be approximated using the quantum dimension.

 It is useful to keep in mind three examples: 1) the Fibonacci anyon $\tau$ with fusion rule $\tau^{\otimes 2}= \mathbf{1}+\tau$ and quantum dimension $d_\tau=\frac{1+\sqrt{5}}{2}$ 2) the Ising anyon $\sigma$ with fusion rules $\sigma^{\otimes 2}=\mathbf{1}+\psi$, $\psi\otimes\sigma=\sigma$, and $\phi^{\otimes 2}=\mathbf{1}$ with  $d_\sigma=\sqrt{2}$ and 3) the $\mbbZ/3$ anyon $\omega$ with fusion rules $\omega^{\otimes 2}=\omega^*$, $\omega\otimes\omega^*=\mathbf{1}$ and $d_\omega=1$.  The ground state degeneracy of Fibonacci anyons in a disk is a Fibonacci number, while the $\mbbZ/3$ has no disk ground state degeneracy and the Ising theory has disk ground state degeneracy of the form $2^n$.

\section{The Role of the Braid Group in TQC}
In this section we will review a number of ways that the braid group plays an essential role in topological quantum computation.   The examples presented here are representative of the author's tastes and are by no means exhaustive.
\subsection{Detecting Non-abelian Anyons via Braiding}
If the braid group representation $\rho_a$ associated with an anyon of type $a$ has abelian image, we say that $a$ is an \emph{abelian} anyon. Abelian anyons have been convincingly demonstrated in the laboratory, and can even be directly shown to admit braidings \cite{nakamura20}.  Notice that anyons that have no ground state degeneracy are automatically abelian: if $\dim(V_{aa}^c)=1$ for a \emph{unique} anyon type $c$ then it can be shown that $\Hom(b,a^{\otimes n})$ is $1$-dimensional for a unique type $b$ for each $n$, and $0$-dimensional for all other types $c\neq b$.  Thus the braid group representation $\rho_a$ described above is $1$ dimensional, hence has abelian image.  In particular if $d_a=1$, then $a$ is an abelian anyon.
 On the other hand, non-abelian anyons are essential if one is to have any interesting (i.e., not phases) braiding gates.  How can one detect if an anyon is non-abelian?  In \cite{RWpra} it is shown that if $a$ is abelian then necessarily $d_a=1$.  The idea is the following where we assume that $a$ is its own anti-particle type for convenience:\footnote{This is not a crucial restriction: a slightly more complicated argument works for all types and in most examples one hopes to construct in the laboratory this is satisfied anyway. }

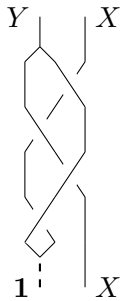
\begin{figure}[th!]
\begin{center}
\begin{tikzpicture}[scale=0.2]%Fig. 2
 \begin{scope}
\begin{scope}[yshift=0cm]
 \draw [thick,dashed] (0,-2)--(0,0);
 \draw (0,0)--(1,1);
 \draw (0,0)--(-1,1);
 \draw (3, -2)--(3,1);
 \draw (0,-2) node[anchor=east] {$\mathbf{1}$};
 \draw (3,-2) node[anchor=west] {${X}$};
\end{scope}
\begin{scope}[yshift=3cm]
 \draw (1,-2)--(-1,1);
 \draw [white, line width=3mm] (-1,-2)--(1,1);
 \draw (-1,-2)--(1,1);
 \draw (3, -2)--(3,1);
\end{scope}
\begin{scope}[yshift=6cm]
 \draw (3,-2)--(1,1);
 \draw [white, line width=3mm] (1,-2)--(3,1);
 \draw (1,-2)--(3,1);
 \draw (-1, -2)--(-1,1);
\end{scope}
\begin{scope}[yshift=9cm]
 \draw (-1,-2)--(1,1);
 \draw [white, line width=3mm] (1,-2)--(-1,1);
 \draw (1,-2)--(-1,1);
 \draw (3, -2)--(3,1);
\end{scope}
\begin{scope}[yshift=12cm]
 \draw (1,-2)--(3,1);
 \draw [white, line width=3mm] (3,-2)--(1,1);
 \draw (3,-2)--(1,1);
 \draw (-1, -2)--(-1,1);
\end{scope}
\begin{scope}[yshift=14cm]
 \draw (0,0)--(0,2);
 \draw (-1,-1)--(0,0);
 \draw (1,-1)--(0,0);
 \draw (3, -1)--(3,2);
 \draw (0,2) node[anchor=east] {${Y}$};
 \draw (3,2) node[anchor=west] {${X}$};
\end{scope}
 \end{scope}
\iffalse \begin{scope}[xshift=8cm,yshift=7cm]
 \draw (0,0) node {$=~~\alpha$};
\end{scope}
\begin{scope}[xshift=13cm,yshift=9cm]
 \draw (0,-4)--(0,0);
 \draw (0,0)--(2,2);
 \draw (0,0)--(-2,2);
 \draw (0,-4) node[anchor=north] {${X}$};
 \draw (-2,2) node[anchor=east] {${Y}$};
 \draw (2,2) node[anchor=west] {${X}$};
\end{scope}
\begin{scope}[xshift=19cm,yshift=7cm]
 \draw (0,0) node {$\neq~~0$};
\end{scope}\fi
\end{tikzpicture}

\caption{A state achieved by creating a pair of $X$ anyons from the vacuum, braiding, and then fusing two $X$ anyons to an anyon $Y$.  This is a non-zero state\label{fig:non-zero state}.}
\end{center}
\end{figure}

\begin{figure}[th!]
\begin{center}
\begin{tikzpicture}[scale=0.25]%Fig. 3
\begin{scope}
%\begin{scope}[yshift=0cm]
 %\draw [thick,dashed] (0,-2)--(0,0);
% \draw (0,0)--(1,1);
% \draw (0,0)--(-1,1);
% \draw (3, -2)--(3,1);
 %\draw (0,-2) node[anchor=east] {$\mathbf{1}$};
 %\draw (3,-2) node[anchor=west] {${X}$};
%\end{scope}
\begin{scope}[yshift=3cm]
 \draw (1,-2)--(-1,1);
 \draw [white, line width=3mm] (-1,-2)--(1,1);
 \draw (-1,-2)--(1,1);
 \draw (3, -2)--(3,1);
\end{scope}
\begin{scope}[yshift=6cm]
 \draw (3,-2)--(1,1);
 \draw [white, line width=3mm] (1,-2)--(3,1);
 \draw (1,-2)--(3,1);
 \draw (-1, -2)--(-1,1);
\end{scope}
\begin{scope}[yshift=9cm]
 \draw (-1,-2)--(1,1);
 \draw [white, line width=3mm] (1,-2)--(-1,1);
 \draw (1,-2)--(-1,1);
 \draw (3, -2)--(3,1);
\end{scope}
\begin{scope}[yshift=12cm]
 \draw (1,-2)--(3,1);
 \draw [white, line width=3mm] (3,-2)--(1,1);
 \draw (3,-2)--(1,1);
 \draw (-1, -2)--(-1,1);
\end{scope}
%\begin{scope}[yshift=14cm]
 %\draw (0,0)--(0,2);
% \draw (-1,-1)--(0,0);
 %\draw (1,-1)--(0,0);
% \draw (3, -1)--(3,2);
% \draw (0,2) node[anchor=east] {${Y}$};
% \draw (3,2) node[anchor=west] {${X}$};
%\end{scope}
 \end{scope}
 \begin{scope}[xshift=8cm,yshift=7cm]
 \draw (0,0) node {$=$};
\end{scope}
\begin{scope}[xshift=12cm,yshift=4cm]
\begin{scope}[yshift=0cm]

\end{scope}
\begin{scope}[yshift=3cm]
 \draw (-1,-4)--(-1,4);
 \draw (1, -4)--(1,4);
 \draw (3, -4)--(3,4);

\end{scope}

\end{scope}
\end{tikzpicture}

\caption{If $X$ is abelian, the sequence of braids has the same effect as the identity braid.\label{fig3}}
\end{center}
\end{figure}
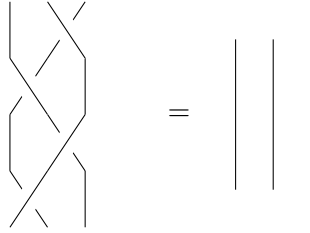

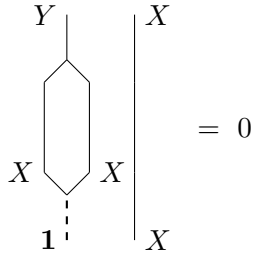
\begin{figure}[th!]
\begin{center}
\begin{tikzpicture}[scale=0.3]%Fig. 3
 
\begin{scope}[xshift=8cm,yshift=7cm]
 %\draw (0,0) node {$~~\gamma$};
\end{scope}
\begin{scope}[xshift=12cm,yshift=4cm]
\begin{scope}[yshift=0cm]
 \draw [thick,dashed] (0,-2)--(0,0);
 \draw (0,0)--(1,1);
 \draw (0,0)--(-1,1);
 \draw (3, -2)--(3,1);
 \draw (0,-2) node[anchor=east] {$\mathbf{1}$};
 \draw (3,-2) node[anchor=west] {${X}$};
\end{scope}
\begin{scope}[yshift=3cm]
 \draw (-1,-2)--(-1,2);
 \draw (1, -2)--(1,2);
 \draw (3, -2)--(3,2);
 \draw (-1,-2) node[anchor=east] {$X$};
  \draw (1,-2) node[anchor=west] {$X$};
\end{scope}
\begin{scope}[yshift=6cm]
 \draw (0,0)--(0,2);
 \draw (-1,-1)--(0,0);
 \draw (1,-1)--(0,0);
 \draw (3, -1)--(3,2);
 \draw (0,2) node[anchor=east] {${Y}$};
 \draw (3,2) node[anchor=west] {${X}$};
\end{scope}
\end{scope}
\begin{scope}[xshift=19cm,yshift=7cm]
 \draw (0,0) node {$=~~0$};
\end{scope}
\end{tikzpicture}

\caption{Since $Y\neq\mathbf{1}$, the state on the left must be zero, and therefore the whole state must be $0$.\label{fig:graph is zero}}
\end{center}
\end{figure}

Suppose that an anyon $X$ of type $a$ has $d_a>1$, yet is abelian.  This implies that there is some anyon $Y$ of non-vacuum type $b\neq\textbf{1}$ so that $\dim(V_{aa}^b)\geq 1$ by a simple eigenvector argument.  We construct a non-zero state in $\Hom(a,a^{\otimes 3})$ by drawing a pair of type $a$ anyons out of the vacuum (we have assumed that the splitting space $\Hom(\mathbf{1},a^{\otimes 2})$ is non-trivial. Next we perform the braiding $\sigma_1\sigma_2\sigma_1^{-1}\sigma_2^{-1}\in\mcB_3$ on this state.  Now $a^{\otimes 2}$ decomposes as a sum of simple objects that includes $b$, so that there is a non-zero vector in $\Hom(a,b\otimes a)$, which can be achieved as a consequence of a fusion of the left two type $a$ anyons to a type $b$ anyon $Y$.  So the process yields the state in Figure \ref{fig:non-zero state}.
Now if the $X$ anyon were abelian, then the image of the braid $\sigma_1\sigma_2\sigma_1^{-1}\sigma_2^{-1}$ must be the identity, as illustrated in figure \ref{fig3}.  Thus, since we cannot create a single anyon out of the vacuum, we obtain the zero state as illustrated in figure \ref{fig:graph is zero}.  This contradicts the assumption that $d_a>1$.

Computing the quantum dimensions of the Fibonacci, Ising and $\mbbZ/3$ anyons we get $\frac{1+\sqrt{5}}{2}$, $\sqrt{2}$ and $1$, respectively.  Thus the Fibonacci and Ising anyons are non-abelian, while the $\mbbZ/3$ anyons are abelian.

\subsection{Braiding Universality}
The foundational paper \cite{FLW02} exhibited the theoretical viability of topological quantum computers by showing that the so-called Fibonacci anyons are universal for quantum computation.  This was achieved by demonstrating that one could efficiently approximately simulate a universal qubit computer on the Fibonacci topological model. Let us briefly recall the qubit model \cite{NC00}.  A qubit is a 2-dimensional vector space $V=\mathbb{C}^2$, typically modeled by some $2$-level quantum system, eg. a spin-system.  One fixes a gate set $\mathcal{G}$ consisting of unitary operators $U_i\in U(V^{\otimes n_i})$.   $N$-qubit quantum circuits are built as compositions of promotions of the $U_i$ to $V^{\otimes N}$, i.e. $I_V^{\otimes a}\otimes U_i\otimes I_V^{\otimes b}$ where $N=a+n_i+b$.  
Each application of a gate is considered a single step, hence the length of the quantum circuit in an algorithm represents consumed time, and is a complexity measure.
A gate set should be physically realizable and complicated enough to perform any computation given enough time.

\begin{definition}
    The gate set $\mathcal{G}$ is said to be \emph{universal} if any unitary operator $X\in U(V^{\otimes N})$ can be efficiently approximated (up to a phase) by a quantum circuit built from $\mathcal{G}$. 
\end{definition}

A typical choice of a universal gate set is
\[ \mc{S} = \{H, \sigma_z^{1/4}, \mathrm{CNOT} \} \]
consisting of the Hadamard, $\pi/8$ and CNOT gates:
\begin{align*}
H &= \frac{1}{\sqrt{2}} \begin{pmatrix}
      1 & 1\\
      1 & -1
    \end{pmatrix}, \quad
\sigma_z^{1/4} = \begin{pmatrix}
      1 & 0\\
      0 & e^{\pi i/4}
    \end{pmatrix}, \quad
\mathrm{CNOT} = \begin{pmatrix}
    1&0&0&0\\
    0&1&0&0\\
    0&0&0&1\\
    0&0&1&0
  \end{pmatrix} 
\end{align*}

Here the efficiency should be that the length of the quantum circuit is a polynomial in the desired accuracy.  The Kitaev-Solovay theorem (see \cite{FLW02} for the most appropriate formulation) states that it is enough to show that the closure of the $N$-qubit circuits in the operator norm topology contains all unitaries in $U(V^{\otimes N})$ up to phases.

For the topological model with gates obtained from braiding as above, the following is the equivalent formulation of universality:

\begin{definition}\label{BraidingUniversal}

An anyon $X$ is called \emph{braiding universal} if, for some $n_0$, the images of $\mcB_n$ on the irreducible sub-representations $W\subset\End(X^{\otimes n})$ are dense in $SU(W)$ for all $n\geq n_0$.
\end{definition}
This is a mathematically precise definition, but in practice it can be quite difficult to verify universality \cite{FLW02}.  The Fibonacci anyon $\tau$, which has quantum dimension $\frac{1+\sqrt{5}}{2}$ is universal, while the Ising anyon $\sigma$ which has quantum dimension $\sqrt{2}$ is not.  In fact the image of the braid group representations associated with the Ising anyon $\sigma$ have \emph{finite} groups as their images--very far from universal!

In practice this appears to be the dichotomy: either an anyon $X$ is braiding universal, or the representations $\rho_X$ have image a finite group, in which case we say $X$ is a \emph{property F} anyon.  In principle, to determine if the image is finite or infinite requires fairly explicit descriptions of the representations, which are not always available. It would be useful to have an indirect way of detecting when an anyon $X$ has associated braid group representations $\rho_X$ with infinite image.  The following gives a very concise conjectural condition:

\begin{conjecture}(\cite{NR11})
An anyon $X$ is  braiding universal if and only $d_x^2\not\in \mathbb{Z}$.
\end{conjecture}
For simplicity we are conflating ``infinite image" with "universality."  But we do not lose too much by doing so: if an irreducible unitary braid group representation $\rho_X:\mcB_n\rightarrow U(V_n)$ has infinite image in $U(V_n)$ then its closure $\overline{\rho_X(\mcB_n)}$ is some infinite compact Lie group $G$.  By analyzing the eigenvalues of the braid group generators' images one finds that $G$ contains, with few exceptions, $SU(V_n),SO(V_n)$ or $Sp(V_n)$ (see \cite{LRW05} for the general method).\footnote{These are the special unitary, special orthogonal and symplectic Lie groups.}  To obtain universality on the nose one hopes for the first case, but one can typically lift the second two cases to special unitaries by taking $n$ slightly larger.

This conjecture has been verified in numerous cases, see eg. \cite{ERW,Ruma,RWenzl,NikGreen,LRW05}.  In particular it is known to be true for anyon models obtained from quantum groups at roots of unity (equivalently, Kac-Moody algebras) as well as all so-called \emph{weakly group-theoretical} braided fusion categories.  This latter class includes the metaplectic anyons of \cite{HNW}.  

\subsection{Measurement Assisted Universality}
Convincing laboratory confirmation of the existence of non-abelian anyons is still a major hurdle.  The simplest model for non-abelian anyons correspond to the Ising categories, known as the Majorana zero modes (or more coloquially, Majorana fermions) \cite{Majorana15}.  It would be a breakthrough to have such a confirmation for the Majorana zero modes.  On the other hand, this could not be used for braiding universal topological quantum computation.  The next simplest non-abelian anyons correspond to the $SU(2)_4$ model (certain metaplectic anyons \cite{HNW}), with some numerical evidence that this is realized by the fractional quantum Hall liquids at filling fraction $\nu=8/3$ \cite{su248thirds}.  But again, these are not braiding universal.  If nature conspires against us (which she so often does) it might be the case that braiding universal anyons are out of reach.  While disappointing, this does not mean that one cannot have topological quantum computers.  In \cite{Cui151,Cui152} protocols for recovering universality from metaplectic anyons are presented.  The idea is to produce a universal gates set from braiding gates and one non-braiding gate achieved through a projective measurement of topological charge.  

\begin{figure}[hbp]
\centering
\begin{picture}(160,60)(-20,-10)
 \label{1-qudit pic}
\put(50,0){\line(0,1){10}}
\put(50,10){\line(1,1){30}}
\put(50,10){\line(-1,1){30}}
\put(70,30){\line(-1,1){10}}
\put(30,30){\line(1,1){10}}

\put(20,42){$X$}
\put(40,42){$X$}
\put(60,42){$X$}
\put(80,42){$X$}
\put(30,19){$a$}
\put(66,19){$b$}
\put(51,-2){$Y$}
\end{picture}\caption{\label{sparseencoding}One qutrit from $SU(2)_4$.}
\end{figure}
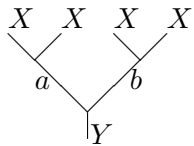

Specifically, the system modeled by the $SU(2)_4$ includes anyons $\textbf{1},Z,X$ and $Y$ with the fusion rules $X^{\otimes 2}=\textbf{1}+Y$ and $Y^{\otimes 2}=\mathbf{1}+Z+Y$, where $Z$ is an abelian anyon satisfying $Z^{\otimes 2}=\textbf{1}$ and $Z\otimes Y=Y$.  A single qutrit is encoded as in figure \ref{sparseencoding}.

The space is $3$ dimensional, corresponding to the pairs $(a,b)$ which can be $(\mathbf{1},Y),(Y,\mathbf{1})$ or $(Y,Y)$.  While braiding the 4 $X$-type anyons provides many gates, braiding alone is not universal.  The additional measurement that must be performed is a projection of the left two $X$-type anyons onto the the vacuum $\mathbf{1}$-type, and the orthogonal complement.  If the left pair total charge is not $\mathbf{1}$ then the right pair total charge is in a coherent superposition of $(Y,\mathbf{1})$ and $(Y,Y)$.  This operation, together with the braiding gates \emph{is} universal.

\subsection{Distinguishing Anyons and Anyon Systems via Braiding}

\begin{figure}[th!]
\begin{center}
\begin{tikzpicture}[scale=0.3]%Fig. 2
\begin{scope}[yshift=5cm]
% \draw (2,2)--(-2,7);
 %\draw [white, line width=3mm] (2,2)--(-2,7);
 %\draw (-2,2)--(2,7);
 \draw (-2,-3)--(2,2);
 \draw [white, line width=3mm] (2,-3)--(-2,2);
 \draw (2,-3)--(-2,2);
 %\draw (-2,-3) node[anchor=east] {${X}$};
 %\draw (2,-3) node[anchor=west] {${X}$};
%  \draw (2,2) node[anchor=west] {${X}$};
 %\draw (-2,2) node[anchor=east] {${X}$};
 %\draw (3, -2)--(3,1);
  \draw (2,2) node[anchor=west] {${X}$};
 \draw (-2,2) node[anchor=east] {${Y}$};
\end{scope}
\begin{scope}[]
% \draw (2,2)--(-2,7);
 %\draw [white, line width=3mm] (2,2)--(-2,7);
 %\draw (-2,2)--(2,7);
 \draw (-2,-3)--(2,2);
 \draw [white, line width=3mm] (2,-3)--(-2,2);
 \draw (2,-3)--(-2,2);
 \draw (-2,-3) node[anchor=east] {${Y}$};
 \draw (2,-3) node[anchor=west] {${X}$};
%  \draw (2,2) node[anchor=west] {${X}$};
 %\draw (-2,2) node[anchor=east] {${X}$};
 %\draw (3, -2)--(3,1);
\end{scope}
\begin{scope}[xshift=10cm]
 \draw (-5,1) node {$=$};
\draw (-2,-3)--(-2,7);
\draw (2,-3)--(2,7);
\draw (-2,-3) node[anchor=east] {${Y}$};
 \draw (2,-3) node[anchor=west] {${X}$};
 \draw (-2,7) node[anchor=east] {${Y}$};
 \draw (2,7) node[anchor=west] {${X}$};
\end{scope}

\end{tikzpicture}

\caption{If the full exchange is trivial for all $Y$, anyon $X$ is \emph{transparent}.\label{fig:transparent}}
\end{center}
\end{figure}
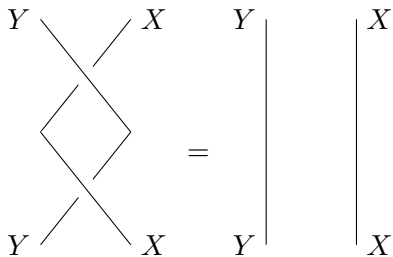

One of the key axioms of a modular category is the \emph{non-degeneracy} of the braiding: this says that the only anyon type $X$ so that the full exchange (double braid, see figure \ref{fig:transparent}) with all anyon types $Y$ is trivial is the vacuum type $\mathbf{1}$.

This turns out to be equivalent to the condition that the matrix with entries $S_{ab}:=\tr(c_{a,b}\circ c_{b,a})$ for all pairs of anyon types $(a,b)$ has $\det(S)\neq 0$ \cite{Brug}.

This implies that, in principle, one may distinguish anyons $X$ and $Y$ by performing all exchanges of $X$ and $Y$ with an anyon of each other type.  One would of course need to perform measurements--in this case the right computation is essentially the one illustrated in figure \ref{fig:TQC}, creating two particle/anti-particle pairs and then projecting onto the vacuum-vacuum state.  These are precisely the entries of the matrix $S$, in diagrams:

\begin{center}\begin{tikzpicture}
\begin{knot}[
    clip width=3,
    flip crossing={2},
    ]
    \strand [ultra thick, red  ] (1.5,0) circle (1.0cm);
    \strand [ultra thick, black] (2.5,0) circle (1.0cm);
    \node (a1) at (1.3,0) [label=$a$]{}; \node (a1) at (2.7,0) [label=$\textcolor{red}{b}$]{};
\end{knot}

\end{tikzpicture}\end{center}
The $S$-matrix entries are the invariants of the Hopf link, with components ``colored" by pairs of anyon types.
Thus the non-degeneracy condition implies we may distinguish anyons by the set of double braidings.

\begin{figure}[th!]
\begin{center}
\begin{tikzpicture}[scale=0.3]%Fig. 2
\begin{scope}[yshift=5cm]
\draw (-3,-1) node[anchor=east] {$c_{X,X}=$};
 \draw (-2,-3)--(2,2);
 \draw [white, line width=3mm] (2,-3)--(-2,2);
 \draw (2,-3)--(-2,2);
 %\draw (-2,-3) node[anchor=east] {${X}$};
 %\draw (2,-3) node[anchor=west] {${X}$};
%  \draw (2,2) node[anchor=west] {${X}$};
 %\draw (-2,2) node[anchor=east] {${X}$};
 %\draw (3, -2)--(3,1);
  \draw (2,2) node[anchor=west] {${X}$};
 \draw (-2,2) node[anchor=east] {${X}$};
\end{scope}

\end{tikzpicture}

\caption{The braiding of $X$ with itself: the trace of this operator has value $d_X\theta_X$.\label{fig:twist}}
\end{center}
\end{figure}
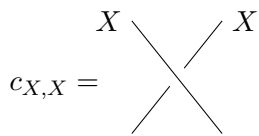

The $S$-matrix records the traces of all double braidings, and hence gives an invariant of the anyon model.  Suppose we wish to pin down precisely which modular category is the model for our anyon system.  What computations/measurements must one perform? 
The quantum dimension $d_a$ of each anyon type appears in the $S$-matrix: it corresponds to setting $b=\mathbf{1}$.  Another invariant is obtained by taking the trace of a single braiding of $X$ with itself, see figure \ref{fig:twist}: this yields the quantity $d_X\theta_X$ for a root of unity $\theta_X$, known as the \emph{topological spin} of $X$.

The $S$-matrix together with the twists $\theta_a$ for all anyon types $a$ form the \emph{modular data} of the theory.  Is the modular data enough to determine the anyon model? The answer turns out to be ``no" \cite{ModularData17}. It is possible for two inequivalent anyon models to have exactly the same modular data: known as \emph{modular isotopes} \cite{isotopes}. By the philosophy that the anyon model should be determined by the topological measurements (mathematically, invariants) one might expect that there is a finite set of measurements that would determine the theory.  That is, one seeks a set of knots and links, including the modular data, so that the invariants one obtains by ``coloring" the components by anyon types completely determines the anyon model.  
 
\begin{figure}[!htb]
\begin{minipage}{0.45\textwidth}
       \centering
        \hfill\begin{tikzpicture}[scale=.5]
 \braid[number of strands=3, rotate=90, style strands={2,3}{red},style strands={1}{blue}] (braid)  a_1 a_2^{-1} a_1 a_2^{-1} a_1;
\end{tikzpicture} 
\end{minipage}
\begin{minipage}{0.5\textwidth}
        \centering
\includegraphics[width=2cm]{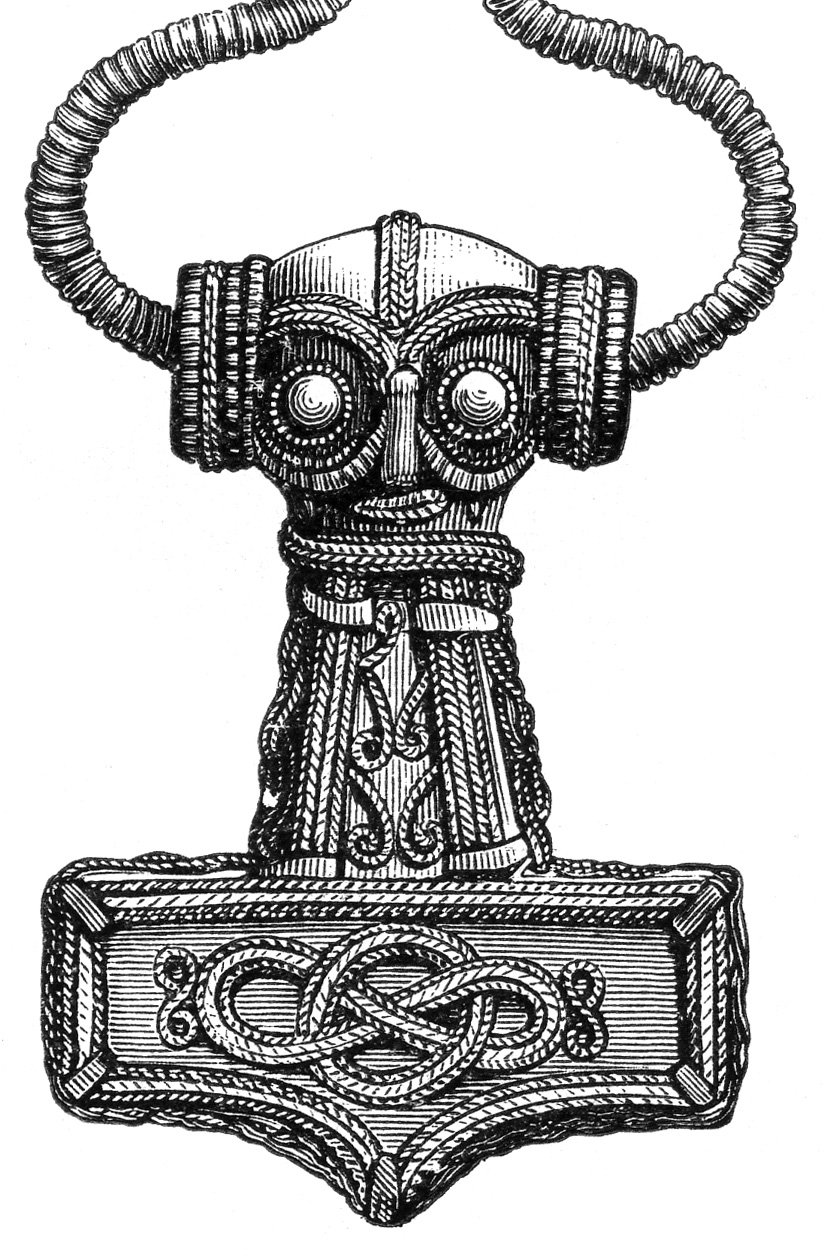}
\end{minipage}
    \caption{Left: A braid with closure the Whitehead link colored to emphasize components. Right: Thor's hammer (Mj\"olnir) artifact featuring the Whitehead link, dating from around 1000 C.E.\label{fig:whitheadbraid}}
\end{figure}
The two component Whitehead link is one candidate (see figure \ref{fig:whitheadbraid}, left) which togehter with the modular data can be used to distinguish some modular isotopes \cite{beyond}.  The Whitehead link is named after the topologist J. C. Whitehead but predates him by about one millenium--it is found on Viking artifacts from around 1000 C.E (figure \ref{fig:whitheadbraid}, right).  Interestingly, the Whitehead link has Gauss linking number $0$.
\section{Beyond Braids}

In the G\=arg\={\i}-Yajnavalkya 
debate one finds the suggestion that the cosmos are braids upon braids upon braids.  Lord Kelvin, in his the vortex-atom theory, proposed that all matter is built from knots.  From our modern perspective these ideas are easily dismissed, but we see glimmers of them in topological phases of matter.  Is there more to this?

\begin{figure}[!htb]
    \centering
\begin{tikzpicture}[line cap=round,line join=round,x=1.0cm,y=1.0cm,scale=.4]
\clip(3.8,5.3) rectangle (17.22,15.04);
\draw [shift={(7.143,8.741)},line width=2.pt]  plot[domain=-0.671:5.135,variable=\t]({1.*0.96*cos(\t r)+0.*0.96*sin(\t r)},{0.*0.96*cos(\t r)+1.*0.96*sin(\t r)});
\draw [shift={(9.600,7.53)},line width=2.pt]  plot[domain=0.030173215557652316:5.92271473679269,variable=\t]({1.*0.94*cos(\t r)+0.*0.94*sin(\t r)},{0.*0.94*cos(\t r)+1.*0.94*sin(\t r)});
\draw [shift={(6.48,11.290319148936172)},line width=2.pt]  plot[domain=-1.3590392941302882:4.500631947720082,variable=\t]({1.*0.95*cos(\t r)+0.*0.95*sin(\t r)},{0.*0.95*cos(\t r)+1.*0.95*sin(\t r)});
\draw [shift={(12.250573813249874,7.8885602503912375)},line width=2.pt]  plot[domain=-5.511828625039063:0.19041787825014977,variable=\t]({1.*0.91*cos(\t r)+0.*0.91*sin(\t r)},{0.*0.91*cos(\t r)+1.*0.91*sin(\t r)});
\draw [shift={(14.252784810126585,9.825601265822785)},line width=2.pt]  plot[domain=-1.8320151706568284:3.834981442045294,variable=\t]({1.*0.98*cos(\t r)+0.*0.98*sin(\t r)},{0.*0.98*cos(\t r)+1.*0.98*sin(\t r)});
\draw [shift={(14.144442413162706,12.623345521023767)},line width=2.pt]  plot[domain=-1.094653172979898:4.566543181398944,variable=\t]({1.*0.99*cos(\t r)+0.*0.99*sin(\t r)},{0.*0.99*cos(\t r)+1.*0.99*sin(\t r)});
\draw [shift={(9.12252737006432,10.210945147042423)},line width=2.pt]  plot[domain=2.4985387663157033:3.3309167550133902,variable=\t]({1.*2.63*cos(\t r)+0.*2.63*sin(\t r)},{0.*2.63*cos(\t r)+1.*2.63*sin(\t r)});
\draw [shift={(9.549538420223378,10.59818426683103)},line width=2.pt]  plot[domain=3.549895026455385:4.359443518328843,variable=\t]({1.*3.16*cos(\t r)+0.*3.16*sin(\t r)},{0.*3.16*cos(\t r)+1.*3.16*sin(\t r)});
\draw [shift={(10.122447072586986,12.345681792593489)},line width=2.pt]  plot[domain=4.473986800724279:4.934235135071265,variable=\t]({1.*5.00*cos(\t r)+0.*5.00*sin(\t r)},{0.*5.00*cos(\t r)+1.*5.00*sin(\t r)});
\draw [shift={(10.730264302656016,11.302785479095485)},line width=2.pt]  plot[domain=4.967087241105105:6.06853738985942,variable=\t]({1.*3.84*cos(\t r)+0.*3.84*sin(\t r)},{0.*3.84*cos(\t r)+1.*3.84*sin(\t r)});
\draw [shift={(11.87801177868133,10.750642452885886)},line width=2.pt] plot[domain=0.14099215481533797:0.9500949355865034,variable=\t]({1.*2.63*cos(\t r)+0.*2.63*sin(\t r)},{0.*2.63*cos(\t r)+1.*2.63*sin(\t r)});
\draw [shift={(10.836581200432741,8.76693850897955)},line width=2.pt]  plot[domain=1.113237616533615:2.3565726019025384,variable=\t]({1.*4.89*cos(\t r)+0.*4.89*sin(\t r)},{0.*4.89*cos(\t r)+1.*4.89*sin(\t r)});
\begin{scriptsize}
\draw [fill=black] (8.22,13.64) circle (2.5pt);
\draw [fill=black] (10.3,14.28) circle (2.5pt);
\draw [fill=black] (12.56,14.06) circle (2.5pt);
\end{scriptsize}
\node (a1) at (8.6,5.3) [label=$n$]{};
\node (a2) at (5.4,7) [label=$n$-{\small 1}]{};
\node (a3) at (4.6,9.9) [label=$n$-{\small 2}]{};
\node (a4) at (13,5.5) [label={\small 1}]{};
\node (a5) at (15.6,8.1) [label={\small 2}]{};
\node (a6) at (15.7,11.5) [label={\small 3}]{};
\end{tikzpicture}

    \caption{The Necklace: motions include leapfrogging and moving each small circle by a rotation of $2\pi/n$ around the center point of the large circle.\label{fig:necklace}}
    
\end{figure}
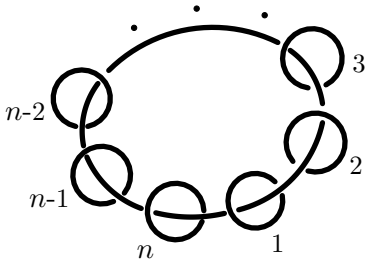

So far we have focused on braids as trajectories of point-like excitations with 2-dimensional ambient space, and have seen that the braid group plays a key role.  Many interesting computational and physical questions can be answered by studying braids.  Can we push this theory to higher dimensions?  Point-like excitations in 3 dimensions are not interesting: they must be bosons or fermions.  But what about other kinds of excitations, such as loop-like vortices? Or, for that matter, knot-like vortices?
At least mathematically, we can describe the dynamics of system in which knots are braided!  The simplest example is known as the \emph{loop braid group}: the motions of $n$ oriented circles in $S^3$.
It has two kinds of generators: the first by ``leapfrogging" the $i$th circle through the $(i+1)$st $\sigma_i$ and the second by loop interchanges $s_i$ for $1\leq i\leq n-1$.  This group is denoted $\mathcal{LB}_n$ and has only been studied relatively recently (see \cite{DamianiSurvey} for a survey). Abstractly, $\mcL\mcB_n$ is the group generated by one copy of the $n$-strand braid group (generated by the $\sigma_i$) and one copy of the symmetric group (generated by the $s_i$) with the additional
(mixed) relations:
\begin{equation*} \sigma_i\sigma_{i+1}s_{i}=
s_{i+1}\sigma_{i}\sigma_{i+1},\quad
 s_is_{i+1}\sigma_{i}=
\sigma_{i+1}s_{i}s_{i+1},\quad 1\leq i\leq n-2, \quad
 \sigma_is_j=s_j\sigma_i \quad
\text{if} \quad |i-j|>1.
\end{equation*} 
Other configurations are possible, such as ``the necklace" as in figure \ref{fig:necklace}.  Representations of the motion group of the necklace, the \emph{necklace braid group} have been studied, for example, in \cite{BKMR}.

\begin{figure}
\includegraphics[width=12cm]{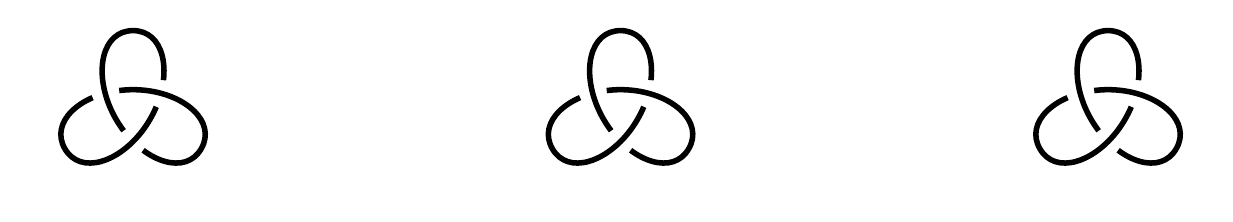}
 \caption{Three trefoils: distinct motions could include passing one such knot through any of the 4 holes in a neighboring knot.\label{fig:threetrefs}}
\end{figure}

A more complicated, but nonetheless intriguing example is the motions of 3 or more trefoil knots as in figure \ref{fig:threetrefs}.  One can imagine many different ``leapfrog'' motions between a pair of such knots. 

Equipped with the mathematical description of these motions, one should be able to approach the theoretical analogues of those described above for 2-dimensional systems in topological phase, such as: \emph{How do we model these systems?}  and \emph{What is the computational power of such a system?}

%\section*{Appendix of Notations and Terminologies}

%\bibliographystyle{amsplain}

%\bibliography{TqcReferences}

%\printbibliography
\bibliographystyle{abbrv}
\bibliography{TqcReferences.bib}
\end{document}